\documentclass[pra,showpacs,eqsecnum,twocolumn]{revtex4}
\usepackage{graphicx}
\usepackage[urlcolor=blue, hyperindex, colorlinks, bookmarks=true,linkcolor=black,citecolor=black]{hyperref}
\begin{document}

\newcommand{\sch}{Schr\"odinger }
\newcommand{\schs}{Schr\"odinger's }
\newcommand{\nn}{\nonumber}
\newcommand{\nl}{\nn \\ &&}
\newcommand{\dg}{^\dagger}
\newcommand{\bra}[1]{\langle{#1}|}
\newcommand{\ket}[1]{|{#1}\rangle}
\newcommand{\psizk}[1]{\ket{\psi_{z}({#1})}}
\newcommand{\psizb}[1]{\bra{\psi_{z}({#1})}}
\newcommand{\erf}[1]{Eq.~(\ref{#1})}
\newcommand{\erfs}[2]{Eqs.~(\ref{#1}) and (\ref{#2})}
\newcommand{\erft}[2]{Eqs.~(\ref{#1}) -- (\ref{#2})}

\title{The non-Markovian stochastic \sch equation for the position unraveling}



\date{\today}
\author{Jay Gambetta}
\affiliation{Centre for Quantum Dynamics, School of Science,
Griffith University, Brisbane 4111, Australia}
\author{H. M. Wiseman} \email{h.wiseman@gu.edu.au}
\affiliation{Centre for Quantum Dynamics, School of Science,
Griffith University, Brisbane 4111, Australia}

\begin{abstract}
An important and well established area of quantum optics is the theory of Markovian stochastic Schr\"odinger equations (or by another name quantum trajectory theory). Recently stochastic Schr\"odinger equations have been developed for non-Markovian systems. In this paper we extend the current known stochastic Schr\"odinger equations for non-Markovian systems to include the position unraveling. We also discuss and illustrate that this stochastic Schr\"odinger equation can have an interpretation under both the orthodox and the de Broglie-Bohm hidden variable interpretation of quantum mechanics. We conclude that only the de Broglie-Bohm hidden variable theory provides a continuous-in-time interpretation of the non-Markovian stochastic Schr\"odinger equation.
\end{abstract}


\maketitle

\section{Introduction}\label{sect:intro}

In nature we are more likely to find a system interacting with an
environment than isolated. Hence a common problem in physics
(especially in the area of quantum optics) is to model open
quantum systems \cite{Car93}. These systems consist of a small
system (the system of interest) immersed in a lager system, which
we refer to as the environment or bath. Due to the large Hilbert
space of the bath it is convenient to describe the system by its
reduced state. The reduced state is defined as
\begin{equation}\label{ReducedState} \rho_{\rm red}(t)={\rm
Tr}_{\rm bath}[\ket{\Psi(t)}\bra{\Psi(t)}],
\end{equation} where $\ket{\Psi(t)}$ is the combined system-bath state, found from
the \sch equation.

It has been shown \cite{Nak58,Zwa60} by a projection-operator
method that we can write a general (non-Markovian) master equation
for the reduced state as
\begin{equation} \label{Master} d_{t}{\rho}_{\rm
red}(t)=-\frac{i}{\hbar}[\hat{H}_{\rm int}(t),\rho_{\rm
red}(t)]+\int_{t_0}^{t}\hat{\cal K}(t,s)\rho_{\rm red}(s) ds,
\end{equation} where $\hat{H}_{\rm int}(t)$ is a system operator
in some interaction picture and $\hat{\cal K}(t,s)$ is the `memory
time' superoperator. It operators on the reduced state
${\rho}_{\rm red}(t)$ and represents how the bath affects the
system. The problem with this equation is that in general the
effect of $\hat{\cal K}(t,s)$ on ${\rho}_{\rm red}(t)$ can not be
explicitly evaluated.

Recently non-Markovian stochastic \sch equations (SSEs)
\cite{DioGisStr98,StrDioGis99,GamWis02a,GamWis02b,Bud00,BasGhi02,Bas03}
have been proposed which allow an alternative procedure for
solving the reduced state. A non-Markovian SSEs is a stochastic
equation for the system state $\psizk{t}$, conditioned on some
noise function $z(t,s)$. (The double time argument in $z$ will be
explained in Sec. \ref{sect: QMT}.) The SSE has the property that
when the outer product of $\psizk{t}$ is averaged over all the
possible $z(t,s)$ one obtains $\rho_{\rm red}$(t). That is,
\begin{equation}\label{Ensemble}
\rho_{\rm red}(t)=E[\psizk{t}\psizb{t}],
\end{equation}
where $E[...]$ denotes an ensemble average over all possible
$z(t,s)$'s.

When using non-Markovian SSEs to solve the reduced state it turns
out that in general we can not explicitly evaluate
$\psizk{t}$\cite{GamWis02b}. That is, we can only explicitly write
a non-Markovian SSE for situations where a reduced state can be
found exactly be other means. However as shown in
Ref.~\cite{YuDioGisStr99} one can perform a post-Markov
perturbation to the non-Markovian SSE, which allows approximate
results for $\rho_{\rm red}(t)$ when the system is close to being
Markovian. In Ref. \cite{GamWis02b} we illustrated a different
perturbation method that allows a perturbative result even when
the system is strongly non-Markovian. However it is only valid for
certain environment correlation functions (memory functions).

Here we are not interested in how to solve current non-Markovian
SSEs. The aim of this paper is to firstly extend the known
non-Markovian unravelings (different functional forms of $z(t,s)$)
to include what we label the `position' unraveling, and then
secondly to outline a non-local hidden variable interpretation of
this non-Markovian SSE. This hidden variable interpretation is
similar to the de Broglie-Bohm interpretation of quantum mechanics
\cite{deB30,Boh52,Hol93}. In other work we will supply a general
hidden variable interpretation to include the coherent and
quadrature unravelings \cite{GamWis03c}. These unravelings are
important to the quantum optics community as they have as their
Markovian limit heterodyne and homodyne detection. The position
unraveling which is presented here does not have a well defined
Markovian limit.

\section{General dynamics for non-Markovian SSEs}
\label{sect: General Dyn}

The aim of this section is to outline the model used to develop
non-Markovian SSEs. The results of this paper are only applicable
to situations when the dynamics of the open quantum system can be
described by the total Hamiltonian
\begin{equation} \label{HamiltonianTotal}
\hat{H}_{\rm sys}(t)\otimes\hat{1}+\hat{1}\otimes\hat{H}_{\rm
bath}+\hat{V},
\end{equation}  where $\hat{H}_{\rm sys}(t)=\hat{H}_{\Omega}+\hat{H}(t)$ is the system
Hamiltonian, and the bath is modelled by a collection of harmonic
oscillators (for optical open quantum systems this corresponds to
the electromagnetic field). In terms of dimensionless position
($\hat{X}_{k}$) and momentum ($\hat{Y}_{k}$) operators, the
Hamiltonian for the bath is
\begin{equation}\label{HamiltonianBath} \hat{H}_{\rm
bath}=\sum_{k}^{K}\frac{\hbar\omega_{k}}{2}(\hat{X}_{k}^{2}+\hat{Y}_{k}^{2}),
\end{equation} where $K$ is the total number of modes in the bath.
These dimensionless operators have the commutator
$[\hat{X}_{j},\hat{Y}_{k}]=i\delta_{j,k}$.

The interaction Hamiltonian, $\hat{V}$ we assume is linear. By
this we mean it has the form
\begin{equation}\label{HamiltonianInteraction2}
\hat{V}=i\hbar\sum_{k}^{K}\Big{[}\hat{L}g^{*}_{k}\frac{(\hat{X}_{k}-i\hat{Y}_{k})}{\sqrt{2}}-\hat{L}\dg
g_{k}\frac{(\hat{X}_{k}+i\hat{Y}_{k})}{\sqrt{2}}\Big{]},
\end{equation}
where $g_{k}$ is the coupling strength of the $k^{\rm th}$ mode to
the system.

For calculation purposes we define an interaction frame such that
the fast dynamics placed on the state by the Hamiltonians
$\hat{H}_{\Omega}$ and $\hat{H}_{\rm bath}$ is moved to the
operators. The unitary operator for this transformation is
\begin{equation}\label{UnitaryFree}
\hat{U}_{0}(t,t_0)=e^{-i(\hat{H}_{\Omega}\otimes\hat{1}+\hat{1}
\otimes\hat{H}_{\rm bath})(t-t_0)/\hbar}.
\end{equation}
This allows us to write the \sch equation as
\begin{equation} \label{IntSchEquation} d_t\ket{{\Psi}(t)} =
-\frac{i}{\hbar}[\hat{H}_{\rm int}(t)+\hat{V}_{\rm int}(t)]
\ket{\Psi(t)},
\end{equation} where $\hat{H}_{\rm int}(t)$ is $H(t)$ in the interaction frame and
\begin{eqnarray} \label{HamiltonianInteractionInt} \hat{V}_{\rm
int}(t)&=& i\hbar\sum_{k}^{K}\Big{[}\hat{L}g_{k}^{*}e^{i\Omega_{k}
(t-t_0)}\frac{(\hat{X}_{k}-i\hat{Y}_{k})}{\sqrt{2}} \nl-\hat{L}\dg
g_{k}e^{-i\Omega_{k}(t-t_0)}\frac{(\hat{X}_{k}+i\hat{Y}_{k})}{\sqrt{2}}
\Big{]},
\end{eqnarray} with $\Omega_{k}=\omega_{k}-\Omega$.  Here we have finally
restricted the form of $\hat{H}_{\Omega}$ to be such that
$\hat{L}$ in the interaction frame simply rotates in the complex
plane at frequency $\Omega$.
 That is $\hat{L}_{\rm int}(t)=
\hat{L}e^{-i\Omega (t-t_0)}$.

\section{Deriving the position non-Markovian SSE}
\label{sect: QMT}

Under the orthodox view \cite{Per93} of quantum mechanics the
quantum state upon measurement undergoes a change which is
consistent with the measurement results. This change in state has
been termed a `collapse'. Whether this collapse is a real physical
event or represents an update in the observer's knowledge will not
be consider here. The central point is that under this view our
observation causes the wavefunction for the state to collapse. The
standard way to model this collapse is to use quantum measurement
theory (QMT) \cite{BraKha92}.

\subsection{Quantum Measurement Theory}

In open quantum systems a measurement is usually performed on the
bath rather than directly on the system. Due to the entanglement
between the bath and the system the measurement on the bath
results in an indirect measurement of the system. For the position
unraveling the measurement performed (on the bath) is to measure
the set of dimensionless position operators $\{\hat{X}_{k}\}$. To
mathematically describe this measurement process we consider the
$K$-observables
\begin{equation}\label{measurement}
  \{{X}_{k}\}=\{(\{x_k\}, \hat\pi_{\{x_k\}}=\ket{\{x_{k}\}}\bra{\{x_{k}\}}\otimes{1}_{\rm sys})\}.
\end{equation} Here the eigenvalues $\{x_{k}\}$ represent
the actual results of the measurement and the eigenstate
$\ket{\{x_{k}\}}$ is the multi-mode state the bath is projected
into upon measurement. Because this theory is done in the
interaction frame the measurement is not exactly a position
measurement. It is position defined in the interaction (rotating)
frame which in the \sch picture (stationary frame) represents a
measurement which cycles between position and momentum.

To find the state of the bath and the system after the measurement
we decompose the projector into measurement operators,
$\hat{M}_{\{x_k\}}$,  by
$\hat{\pi}_{\{x_{k}\}}=\hat{M}_{\{x_k\}}\dg\hat{M}_{\{x_k\}}$.
With this measurement operator the combined state after a
measurement at time $t$, which yielded results $\{x_{k}\}$ is
\begin{equation}\label{Combined}
  \ket{\Psi_{\{x_{k}\}}(t)}=\frac{\hat{M}_{\{x_{k}\}}\ket{\Psi(t)}}{\sqrt{P(\{x_{k}\},t)}},
\end{equation} where $P(\{x_{k}\},t)$ is the probability density and is defined by
\begin{equation}\label{probability}
  P(\{x_{k}\},t)=\bra{\Psi(t)}\hat{\pi}_{\{x_{k}\}}\otimes\hat{1}_{\rm sys}\ket{\Psi(t)}.
\end{equation}

For the rank-one projector, $\ket{\{x_{k}\}}\bra{\{x_{k}\}}$, the
measurement operator which satisfies the above decomposition is
$\hat{M}_{\{x_k\}}=\ket{\{n_k\}}\bra{\{x_k\}}$, where
$\ket{\{n_{k}\}}$ is arbitrary and is the state the bath is left
in after the measurement. For most optical situations this will be
a vacuum state $\ket{\{0_{k}\}}$ (as in optical measurement the
detector usually absorbs the photon). This results in the combined
state becoming
$\ket{\Psi_{\{x_{k}\}}(t)}=\ket{\{0_k\}}\ket{\psi_{\{x_{k}\}}(t)}
$, where $\ket{\psi_{\{x_{k}\}}(t)}$ is defined as
\begin{equation}\label{Conditioned}
  \ket{\psi_{\{x_{k}\}}(t)}=\frac{\langle\{x_{k}\}\ket{\Psi(t)}}{\sqrt{P(\{x_{k}\},t)}},
\end{equation}
which we label a conditioned system state. This conditioned system
state has the property that
\begin{equation}
  \rho_{\rm
  red}(t)=\int
  {P(\{x_{k}\},t)}\ket{\psi_{\{x_{k}\}}(t)}\bra{\psi_{\{x_{k}\}}(t)}d\{x_{k}\}.
\end{equation}

If we consider $\{x_{k}\}$ to be a set of random variables
$\{x_{k}(t)\}$ chosen from the distribution $ {P(\{x_{k}\},t)}$
then we can rewrite the reduced state as
\begin{equation}\label{AverageReduced}
  \rho_{\rm
  red}(t)=E[\ket{\psi_{\{x_{k}(t)\}}(t)}\bra{\psi_{\{x_{k}(t)\}}(t)}],
\end{equation} where $E$ denotes an average over $
{P(\{x_{k}\},t)}$. This is the requirement outlined in Sec.
\ref{sect:intro} for a solution to a SSE. This suggests that the
time derivative of $\ket{\psi_{\{x_{k}(t)\}}(t)}$ will be a
non-Markovian SSE. To calculate this we need to be able to
generate a self consistent differential equation for
$\ket{\psi_{\{x_{k}(t)\}}(t)}$. This is complicated and to
simplify this procedure we introduce linear quantum measurement
theory (LQMT).

\subsection{Linear Quantum Measurement Theory}

LQMT uses the same principles as QMT except we use an ostensible
distribution, $\Lambda(\{x_{k}\})$, in place of the actual
distribution \cite{GoeGra94,Wis96}. Using this distribution the
linear conditional state is
\begin{equation}\label{LinearConditioned}
  \ket{\bar{\psi}_{\{x_{k}\}}(t)}=\frac{\langle\{x_{k}\}\ket{\Psi(t)}}{\sqrt{\Lambda(\{x_{k}\})}}.
\end{equation} The bar above this linear state signifies that the state is
unnormalised. As before the reduced state can be written as
\begin{equation}
  \rho_{\rm
  red}(t)=\int
  {\Lambda(\{x_{k}\})}\ket{\bar{\psi}_{\{x_{k}\}}(t)}\bra{\bar{\psi}_{\{x_{k}\}}(t)}d\{x_{k}\}.
\end{equation} Thus $\rho_{\rm
red}(t)=\bar{E}[\ket{\bar{\psi}_{\{x_{k}(t)\}}(t)}\bra{\bar{\psi}_{\{x_{k}(t)\}}(t)}]$
where $\bar{E}$ denotes $\{x_{k}(t)\}$ is chosen from the
ostensible distribution $\Lambda(\{x_{k}\})$. Because
$\Lambda(\{x_{k}\})$ is time independent
$\{x_{k}(t)\}=\{x_{k}(t_0)\}$ for all $t$ and $d_{t}
\ket{\bar{\psi}_{\{x_{k}\}}(t)}=
\partial_{t}\ket{\bar{\psi}_{\{x_{k}\}}(t)}$.

Using \erf{LinearConditioned} the time derivative of
$\ket{\bar{\psi}_{\{x_{k}\}}(t)}$ becomes
\begin{equation}
\partial_{t}\ket{\bar{\psi}_{\{x_{k}\}}(t)}
=\frac{\langle\{x_{k}\}|d_{t}\ket{\Psi(t)}}{\sqrt{\Lambda(\{x_{k}\})}}.
\end{equation} Substituting into this
\erf{IntSchEquation} and the Hamiltonian defined in
\erf{HamiltonianInteractionInt} the differential equation for the
linear system state becomes
\begin{widetext}\begin{eqnarray}\label{linearSSE1a}
\hspace{-.5cm}\partial_{t}\ket{\bar{\psi}_{\{x_{k}\}}(t)}&=&
\Big{[}-\frac{i}{\hbar}\hat{H}_{\rm
int}(t)+\frac{\sum_{k}(\hat{L}g_{k}^{*}x_{k}e^{i\Omega_{k}(t-t_0)}-\hat{L}\dg
g_{k}x_{k}e^{-i\Omega_{k}(t-t_0)}
}{\sqrt{2}}\Big{]}\ket{\bar{\psi}_{\{x_{k}\}}(t)} \nl-\frac
{\sum_{k}(\hat{L}g_{k}^{*}e^{i\Omega_{k}(t-t_0)} +\hat{L}\dg
g_{k}e^{-i\Omega_{k}(t-t_0)})}{\sqrt{2}}\frac{\partial_{x_{k}}\langle\{x_{k}\}\ket{\Psi(t)}}{\sqrt{\Lambda(\{x_{k}\})}}.
\end{eqnarray}
Choosing the ostensible distribution to be
\begin{equation}\label{Ostensible}
  \Lambda(\{x_{k}\})=|\bra{\{x_{k}\}}\{0_{k}\}\rangle|^{2}
  =\prod_{k}\frac{\exp(-x_{k}^{2})}{\sqrt{\pi}},
\end{equation} results in
\begin{eqnarray}\label{linearSSE1}
\hspace{-.5cm}\partial_{t}\ket{\bar{\psi}_{\{x_{k}\}}(t)}&=&\Big{[}-\frac{i}{\hbar}\hat{H}_{\rm
int}(t)+\sum_{k}\hat{L}g_{k}^{*}x_{k}\sqrt{2}e^{i\Omega_{k}(t-t_0)}-\frac
{\sum_{k}(\hat{L}g_{k}^{*}e^{i\Omega_{k}(t-t_0)} +\hat{L}\dg
g_{k}e^{-i\Omega_{k}(t-t_0)})}{\sqrt{2}}\partial_{x_{k}}\Big{]}\ket{\bar{\psi}_{\{x_{k}\}}(t)}.\nl
\end{eqnarray} \end{widetext} Note this ostensible distribution was chosen for simplicity (it is equal to
the real distribution at $t=t_0$).

To define a SSE from \erf{linearSSE1} we make the random variables
substitution $x_{k}\rightarrow x_{k}(t)$. This transforms the
linear state as
$\ket{\bar{\psi}_{\{x_{k}\}}(t)}\rightarrow\ket{\bar{\psi}_{\{x_{k}(t)\}}(t)}$.
For the linear SSE the random variable $x_{k}(t)$ must satisfy
$\bar{E}[x_{k}(t)]=0$ and
$\bar{E}[x_{k}(t)x_{l}(t)]=\delta_{k,l}/2$ for all $t$ (due to
\erf{Ostensible}). It can be still argued that, even with this
substitution, this is not strictly a SSE as to solve this we must
simultaneously solve for all possible $\{x_{k}(t)\}$ (due to the
partial derivative in \erf{linearSSE1}). To make it a SSE we need
to replace the derivatives by operators, but before we do this, to
keep with past literature, it is convenient to define a noise
function $z(t,s)$ as
\begin{eqnarray}\label{z}
  z(t,s)&=&\sum_{k}g_{k}\sqrt{2}x_{k}(t)e^{-i\Omega_{k}(s-t_0)},
 \end{eqnarray} With this definition we can define the following two bath-correlation
function \begin{widetext}
\begin{eqnarray}\label{coorel1}
  \bar{E}[z(t,s)z^{*}(t,s')]&=&\sum_{k,l}g_{k}g_{l}^{*}2e^{-i\Omega_{k}(s-t_0)+i\Omega_{k}(s'-t_0)}\bar{E}[x_{k}(t)x_{l}(t)]\\ &=&\sum_{k}|g_{k}|^{2}e^{-i\Omega_{k}(s-'s)}=\alpha(s-s'),  \\
\bar{E}[z(t,s)z(t,s')]&=&\sum_{k}g_{k}^{2}e^{-i\Omega_{k}(s+s'-2t_0)}=\gamma(s+s')
\end{eqnarray} and the $K$ partial derivatives become
\begin{eqnarray}\label{partialToFunctional}
  \partial_{x_{k}}&=&\int_{t_0}^{t}\Big{[}\delta_{z(t,s)}\frac{\delta z(t,s)}
  {\delta x_{k}(t)}ds+\delta_{z^{*}(t,s)}\frac{\delta z^{*}(t,s)}
  {\delta x_{k}(t)}ds\Big{]}\nn\\&=&\int_{t_0}^{t}\sqrt{2}\Big{[}\delta_{z(t,s)}
  g_{k}e^{-i\Omega_{k}(s-t_0)}ds+\delta_{
  z^{*}(s)}g^{*}_{k} e^{i\Omega_{k}(s-t_0)}ds\Big{]},
\end{eqnarray} where $\delta_{z^{*}(t,s)}$ and $\delta_{z^{*}(t,s)}$
are functional derivatives. In terms of the noise function,
\erf{linearSSE1} becomes
\begin{eqnarray}\label{linearSSE2a}
\partial_{t}\ket{\bar{\psi}_{z}(t)}&=&\Big{\{}-\frac{i}{\hbar}\hat{H}_{\rm
int}(t)+\hat{L}z^{*}(t,t)-\hat{L}\int_{t_0}^{t}\Big{[}\alpha^{*}(t-s)\delta_{
z(t,s) }+\gamma^{*}(t+s)\delta_{z^{*}(t,s)}
\Big{]}ds-\hat{L}\dg\int_{t_0}^{t}\Big{[}\alpha(t-s)\delta_{
z^{*}(t,s) }\nl+\gamma(t+s)\delta_{z(t,s)} \Big{]}ds\Big{\}}
\ket{\bar{\psi}_{z}(t)}.
\end{eqnarray} \end{widetext}Assuming the initial combined state to be
$\ket{\Psi(t_0)}=\ket{\{0\}_k}\ket{\psi(t_0)}$ it is easily shown
that $\ket{\bar{\psi}_{z}(t_0)}\equiv\ket{{\psi}(t_0)}$ thus the
functional derivative with respect to $z(t,s)$ in the above
equation will always have zero contribution. That is, we can
rewrite \erf{linearSSE2a} as
\begin{eqnarray}\label{linearSSE2}
\partial_{t}\ket{\bar{\psi}_{z}(t)}&=&\Big{\{}-\frac{i}{\hbar}\hat{H}_{\rm
int}(t)+\hat{L}z^{*}(t,t)-\int_{t_0}^{t}\Big{[}\hat{L}\gamma^{*}(t+s)\nl+\hat{L}\dg\alpha(t-s)\Big{]}\delta_{z^{*}(t,s)}
ds\Big{\}} \ket{\bar{\psi}_{z}(t)}.
\end{eqnarray} This equation appears nicer than \erf{linearSSE1}, but it is
essentially the same equation and we still have the problem of
representing the functional derivative by operators. To do this we
make the two ansatzen,
\begin{eqnarray}\label{Ansatz}
\hat{A}_{z}(t)\ket{\bar{\psi}_{z}(t)}&=&\int_{0}^{t}ds\alpha(t-s)\delta_{z^{*}(t,s)}\ket{\bar{\psi}_{z}(t)},\\
\label{Ansatz2}
\hat{B}_{z}(t)\ket{\bar{\psi}_{z}(t)}&=&\int_{0}^{t}ds\gamma^{*}(t+s)\delta_{z^{*}(t,s)}\ket{\bar{\psi}_{z}(t)}.
\end{eqnarray}
With these ansatzen the linear non-Markovian SSE for the position
unraveling is
\begin{eqnarray}\label{linearSSE3}
\partial_{t}\ket{\bar{\psi}_{z}(t)}&=&\Big{\{}-\frac{i}{\hbar}\hat{H}_{\rm
int}(t)+\hat{L}z^{*}(t,t)-\hat{L}\hat{B}_{z}(t)\nl
-\hat{L}\dg\hat{A}_{z}(t)\Big{\}} \ket{\bar{\psi}_{z}(t)}.
\end{eqnarray} In general these two ansatzen can not be solved,
but as in coherent and quadrature unravelings
\cite{GamWis02b,YuDioGisStr99} there are perturbation techniques
which could be extended to include this unraveling and would allow
a perturbative solution. This we leave for later work.

\subsection{The Non-Markovian SSE}

In the above section we have presented the linear SSE for the
position unraveling. To generate the actual non-Markovian SSE from
this linear equation we have to first of all generate the set of
random variable $\{x_{k}(t)\}$, which obey the real distribution
$P(\{x_{k}\},t)$. These in general will be time dependent. To do
this we use a Girsanov transformation \cite{GatGis91,DioGisStr98}
to relate the real probability distribution with the ostensible
one, that is
\begin{equation}\label{Girsanov}
  P(\{x_{k}\},t)=\Lambda(\{x_{k}\})\langle{\bar{\psi}_{\{x_{k}\}}(t)}
  \ket{\bar{\psi}_{\{x_{k}\}}(t)}.
\end{equation} Taking the time derivative of this and using \erf{linearSSE1} gives (after some
manipulating)
\begin{eqnarray}\label{Fokker}
  d_{t}P(\{x_{k}\},t)&=&-\sum_{k}\partial_{x_{k}}\Big{\{}\bra{{\psi}_{\{x_{k}\}}(t)}[\hat{L}g_{k}^{*}
  e^{i\Omega_{k}(t-t_0)} \nl +\hat{L}\dg g_{k} e^{-i\Omega_{k}
  (t-t_0)}]\ket{{\psi}_{\{x_{k}\}}(t)}\nl\times P(\{x_{k}\},t)\Big{\}}/\sqrt{2}.
  \end{eqnarray}
This is effectively a drift equation for the probability density.
It has associated with it the following set of differential
equations
\begin{equation}\label{SDE}
  d_{t} x_{k}(t)={[\langle\hat{L}\rangle_{t}g_{k}^{*}
  e^{i\Omega_{k}(t-t_0)} +\langle\hat{L}\dg\rangle_{t} g_{k} e^{-i\Omega_{k}
  (t-t_0)}]}/{\sqrt{2}}
\end{equation}
where $\langle
\hat{L}\rangle_{t}=\bra{{\psi}_{\{x_{k}(t)\}}(t)}\hat{L}\ket{{\psi}_{\{x_{k}(t)\}}(t)}$.
Integrating this gives
\begin{eqnarray}\label{xt}
 x_{k}(t)&=&x_{k}(t_0)+\int_{t_0}^{t} dt'
[\langle\hat{L}\rangle_{t'}g_{k}^{*}
  e^{i\Omega_{k}(t'-t_0)} \nl +\langle\hat{L}\dg\rangle_{t'} g_{k} e^{-i\Omega_{k}
  (t'-t_0)}]/{\sqrt{2}},
\end{eqnarray} where $x_{k}(t_0)$ is the random variable associated
with the distribution
$\bra{\Psi(t_0)}\hat{\pi}_{x_{k}}\ket{\Psi(t_0)}$, which for an
initial combined state of the form
$\ket{\Psi(t_0)}=\ket{\{0_k\}}\ket{\psi(t_0)}$ is equivalent to
the above ostensible distribution, \erf{Ostensible}. The noise
function for the real distribution, defined in \erf{z}, becomes
\begin{equation}\label{zactual}
  z(t,s)=z(t_0,s)+\int_{t_0}^{t} dt'
{[\langle\hat{L}\rangle_{t'}\alpha(s-t')
+\langle\hat{L}\dg\rangle_{t'} \gamma(s+t')]}
\end{equation} where $z(t_0,s)$ is equivalent to the noise function used
in the linear non-Markovian SSE.

To generate the non-Markovian SSE we use
\erfs{Conditioned}{LinearConditioned} to rewrite the conditioned
state
$\ket{{\psi}_{\{x_{k}(t)\}}(t)}=\ket{{\psi}_{\{x_{k}\}}(t)}|_{\{x_{k}=x_{k}(t)\}}$
as
\begin{equation}
\ket{{\psi}_{\{x_{k}(t)\}}(t)}=\frac{\ket{\bar{\psi}_{\{x_{k}(t)\}}(t)}}
{\sqrt{\langle\bar{\psi}_{\{x_{k}(t)\}}(t)\ket{\bar{\psi}_{\{x_{k}(t)\}}(t)}}}.
\end{equation}
As long as $\{x_{k}(t)\}$ in $\ket{\bar{\psi}_{\{x_{k}(t)\}}(t)}$
are the random variables associated with the real distribution,
\erf{xt}, this will obey the correct statistics defined in
\erf{AverageReduced}. Using \erf{linearSSE3} and the standard
definition for a total time derivative gives the following SSE
(after some manipulation)\begin{widetext}
\begin{eqnarray}\label{SSE2}
d_{t}\ket{{\psi}_{z}(t)}&=&\Big{\{}-\frac{i}{\hbar}\hat{H}_{\rm
int}(t)+(\hat{L}-\langle\hat{L}\rangle_{t})z^{*}(t)-(\hat{L}-\langle\hat{L}\rangle_{t})
\hat{B}_{z}(t)+\langle(\hat{L}-\langle\hat{L}\rangle_{t})
\hat{B}_{z}(t)\rangle_{t}
-(\hat{L}\dg-\langle\hat{L}\dg\rangle_{t})\hat{A}_{z}(t)
\nl+\langle(\hat{L}\dg-\langle\hat{L}\dg\rangle_{t})\hat{A}_{z}(t)\rangle_{t}
\Big{\}} \ket{{\psi}_{z}(t)}, 
\end{eqnarray} \end{widetext} where
$\ket{{\psi}_{z}(t)}\equiv\ket{{\psi}_{\{x_{k}(t)\}}(t)}$. Here we
have again used the above ansatzen to remove the partial
derivatives with respect to $\{x_{k }\}$. Under the orthodox
interpretation, the solution of this equation at time $t$ is the
state of the system given that a measurement has been performed on
the bath at that time and yielded results $\{x_{k}(t)\}$. Since
this measurement would change the bath state, the future evolution
of the system would not be the same as if the measurement had not
been performed. In other words, the solutions at different times
correspond to different physical situation. Thus the linking of
these solutions to make a trajectory for the system state is a
fiction.

\section{SIMPLE APPLICATION}
\label{sect: simple}

\begin{figure}
\begin{center}
\includegraphics[width=.45\textwidth]{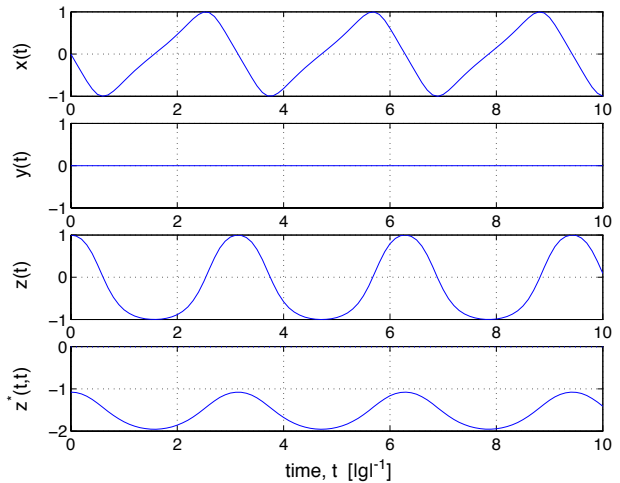}
\end{center}
\caption{\label{Fig.1} An example trajectory for the position
non-Markovian SSE for a TLA in a single mode bath. Also shown is
the real (solid) and imaginary (dotted) part of the noise
function. Note the imaginary part for  this system is equal to
zero for all time. All calculations where done numerically with
$g=1$, a time step size of 0.0001 and an excited state initial
condition.}
\end{figure}

\begin{figure}\begin{center}
\includegraphics[width=.45\textwidth]{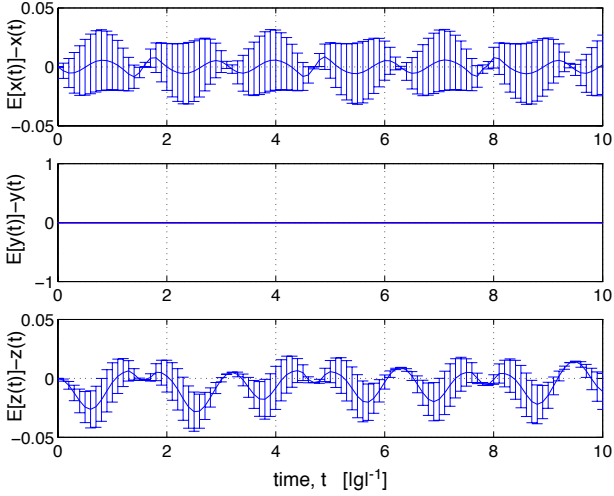}
\end{center}
\caption{\label{Fig.2} The difference between the ensemble average
of 1000 trajectories for the position unraveling and the master
equation for a TLA in a single mode bath. Other details are as in
figure \ref{Fig.1}.}
\end{figure}

To illustrate that this is a correct unraveling for a
non-Markovian open quantum system, in this section we apply this
theory to a simple model. The model is a two level atom (TLA)
coupled linearly to a single mode bath ($K=1$) with no detuning.
To get an exact solution for comparison we first of all solve the
total \sch equation for the combined state. For this system the
\sch equation in the interaction frame is
\begin{equation}\label{schmodel}
  \ket{\dot{\Psi}(t)}=(g^{*}\hat{\sigma}\hat{a}\dg-g\hat{\sigma}\dg\hat{a})\ket{\Psi(t)}
\end{equation} where $\hat{L}=\hat{\sigma}=\ket{b}\bra{e}$ is the lowering operator for the
TLA and $\ket{e}$ and $\ket{b}$ are the excited and ground states
of the TLA. The combined state can be written in terms of photon
number states and atomic states as
\begin{equation}
  \ket{\Psi(t)}=\sum_{s=e,b}\sum_{n}c_{s,n}\ket{s}\ket{n}.
\end{equation} Substituting this into \erf{schmodel} and using the fact that since the
bath state is initially in a vacuum state the only nonzero
amplitudes are $c_{e,0}$, $c_{b,0}$ and $c_{b,1}$, we get the
following solutions (with $t_0=0$)
\begin{eqnarray}
c_{e,0}(t)&=&c_{e,0}(0)\cos[|g|(t-t_0)],\\
c_{b,0}(t)&=&c_{b,0}(0),\\
c_{b,1}(t)&=&c_{e,0}(0)\sin[|g|(t-t_0)]e^{-i\theta},
\end{eqnarray} where $\theta$ is the argument of the complex coupling
constant $g$. Thus the reduced state is simply
\begin{widetext}\begin{eqnarray} \label{ReduceStateModel}
  \rho_{\rm
  red}(t)&=&c^{2}_{e,o}(t_0)\cos^{2}[|g|(t-t_0)]\ket{e}\bra{e}
  +\{c^{2}_{b,0}(t_0)+c^{2}_{e,o}(t_0)\sin^{2}[|g|(t-t_0)]\}\ket{b}\bra{b}\nl+
  [c_{e,o}(t_0)c^{*}_{b,0}(t_0)\ket{e}\bra{b}+
  c^{*}_{e,o}(t_0)c_{b,0}(t_0)\ket{b}\bra{e}]\cos[|g|(t-t_0)].
\end{eqnarray}

The general non-Markovian SSE defined in \erf{SSE2} for this
simple system becomes
\begin{eqnarray}\label{SSE3}
d_{t}\ket{{\psi}_{z}(t)}&=&\Big{\{}(\hat{\sigma}-\langle\hat{\sigma}\rangle_{t})
z^{*}(t)-(\hat{\sigma}-\langle\hat{\sigma}\rangle_{t})
\hat{B}_{z}(t)+\langle(\hat{\sigma}-\langle\hat{\sigma}\rangle_{t})
\hat{B}_{z}(t)\rangle_{t}
-(\hat{\sigma}\dg-\langle\hat{\sigma}\dg\rangle_{t})\hat{A}_{z}(t)\nl
+\langle(\hat{\sigma}\dg-\langle\hat{\sigma}\dg\rangle_{t})\hat{A}_{z}(t)\rangle_{t}
\Big{\}} \ket{{\psi}_{z}(t)},
\end{eqnarray} \end{widetext} where
\begin{equation}
  z(t)=g\sqrt{2}x_1(t)
\end{equation}and
\begin{equation}
  x_1(t)=x_1(t_0)+\int_{t_0}^{t}dt'(g^{*}\langle\hat\sigma\rangle_{t'}+g\langle\hat\sigma
\dg\rangle_{t'})/\sqrt{2}.
\end{equation} The value of the random variable $x_1(t_0)$ is determined by the initial distribution
\begin{equation}\label{lambdaexample}
 P(x_1,t_0)=\Lambda(x_1)=\frac{\exp(-x_1^{2})}{\sqrt{\pi}}.
\end{equation}
To find functional form of the two operators $\hat{A}_{z}(t)$ and
$\hat{B}_{z}(t)$ we use \erfs{Ansatz}{Ansatz2} and assume
\begin{eqnarray} \label{ansatzsystem}
\delta_{z^{*}(s)}\ket{\bar{\psi}_{z}(t)}&=&f(t,s)\hat{\sigma}
\ket{\bar{\psi}_{z}(t)},
\end{eqnarray}
This results in $\hat{A}_{z}(t)={A}_{z}(t)\hat{\sigma}$, and
$\hat{B}_{z}(t)={B}_{z}(t)\hat{\sigma}$, where
\begin{eqnarray}
{A}_{z}(t)&=&\int_{t_0}^{t}ds|g|^{2}f(t,s),\\
{B}_{z}(t)&=&\int_{t_0}^{t}ds {g^{*}}^{2}f(t,s).
\end{eqnarray} That is ${A}_{z}(t)$ and ${B}_{z}(t)$ are independent of the noise function for this example.
 Taking the time
derivative of these equations gives 
\begin{eqnarray}\label{sub1}
\dot{A}_{z}(t)&=&|g|^{2}+\int_{t_0}^{t}|g|^{2}\dot{f}(t,s)ds,\\
\label{sub2}
\dot{B}_{z}(t)&=&{g^{*}}^{2}+\int_{t_0}^{t}{g^{*}}^{2}\dot{f}(t,s)ds.
\end{eqnarray}  Here we have used the fact that
$f(t,t)=1$ \cite{GamWis02a}. To find the form of $\dot{f}(t,s)$ we
firstly need to define the linear non-Markovian SSE for this
system. It is
\begin{equation} \label{linearsystem}
  \partial_t
  \ket{\bar{\psi}_{z}(t)}=[\hat{\sigma}z^{*}(t)-{A}_{z}(t)\hat{\sigma}\dg\hat{\sigma}]\ket{\bar{\psi}_{z}(t)},
\end{equation} where $z(t)=g\sqrt{2}x_1(t)$ with $x_1(t)$ in the
linear case being defined to have a value given by $\Lambda(x_1)$
[see \erf{lambdaexample}]. Secondly we use the following
consistency conditions
\begin{eqnarray}
  \delta_{z^{*}(s)}\partial_{t}\ket{\bar{\psi}_{z}(t)}&=&
\partial_{t}\delta_{z^{*}(s)}
  \ket{\bar{\psi}_{z}(t)}.
\end{eqnarray} Substituting \erf{linearsystem} and
\erf{ansatzsystem} into this equation gives the following
differential equation
\begin{eqnarray}
\dot{f}(t,s)&=&f(t,s){A}_{z}(t),
\end{eqnarray}
which when substituted into \erfs{sub1}{sub2} gives 
\begin{eqnarray}
\dot{A}_{z}(t)&=&|g|^{2}+{A}^{2}_{z}(t),\\
\dot{B}_{z}(t)&=&{g^{*}}^{2}+{B}_{z}(t){A}_{z}(t).
\end{eqnarray}
Solving this set of coupled differential equation with the initial
conditions ${A}_{z}(t_0)={B}_{z}(t_0)=0$ gives
${A}_{z}(t)=|g|\tan[|g|(t-t_0)]$, and
${B}_{z}(t)=e^{-i2\theta}|g|\tan[|g|(t-t_0)]$. Thus \erf{SSE3} is
now numerically solvable. An example solution for
$\ket{\psi_{z}(t)}$ is shown in Fig \ref{Fig.1}. Here it is
observed that unlike a Markov trajectory the evolution is smooth.
This is because with only one bath mode the bath correlation time
is non-zero.
 In this
figure the quantum state is displayed in the Block representation,
that is
$x(t)=\langle\hat\sigma\rangle_t+\langle\hat\sigma\dg\rangle_t$,
$y(t)=-i\langle\hat\sigma\rangle_t+i\langle\hat\sigma\dg\rangle_t$
and
$z(t)=\langle\hat\sigma\dg\hat\sigma\rangle_t-\langle\hat\sigma\hat\sigma\dg\rangle_t$.
To show that this equation does reproduce the reduced state, the
difference between the ensemble average of 1000 trajectories and
the reduced state, \erf{ReduceStateModel}, is shown in Fig.
\ref{Fig.2}.  It is observed that the ensemble averages does
reproduce the reduced state within statistical error.

\section{A HIDDEN VARIABLE INTERPRETATION OF THE POSITION NON-MARKOVIAN
SSE} \label{sect: hidden}

In this section we present an alternative interpretation of the
position non-Markovian SSE. The hidden variable interpretation we
use is similar to the de Broglie-Bohm interpretation. The only
difference is our position variables are `positions' in a
reference frame which rotates in phase space (the interaction
frame). This results in our trajectories for the positions to be
different from the standard de Broglie-Bohm trajectories
\cite{deB30,Boh52,Hol93}.  Moving to this frame simply means that
the position states as we have defined them actual correspond to,
in a stationary frame, a state which rotates between position and
momentum. This is not a problem provided the \sch equation used to
develop the trajectories is in the interaction (rotating) frame.

In the standard de Broglie-Bohm theory, position has a reality
prior to the measurement and its trajectory is deterministic
assuming its initial position is known. To account for the
weirdness of quantum mechanics its trajectory can be very
non-classical. Specifically, the trajectory is influence by a
extra potential, the quantum potential \cite{Boh52,Hol93}. This
quantum potential depends on the solution of the \sch equation,
and is why the wavefunction is called the guiding wave
\cite{deB30,Boh52,Hol93}. Thus under this interpretation the
wavefunction is a real entity.

In this paper we will not be introducing the quantum potential as
we can describe the trajectories with reference to only the
wavefunction as a guiding wave.  To mathematically describe this
interpretation we use the fact that since the probability density
is continuous in time and configuration space
$\{x_{k}~\epsilon~\Re^{K}\}$ and is a conserved quantity, it must
obey a continuity equation,
\begin{equation}\label{Continuity}
  \partial_{t} P(\{x_{j}\},t)=-\sum_{k}\partial_{x_{k}}j_{k}(\{\
  x_{j}\},t),
\end{equation} where $j_{k}(\{x_{j}\},t)$ is the current
density and is related to the velocity field by $j_{k}(\{
x_{j}\},t)=P(\{x_{j}\},t){\rm v}_{k}(\{x_{j}\},t)$. Each
individual trajectory is then found by
\begin{equation}\label{VelocityA}
d_{t}{x}_{k}(t)={\rm v}_{k}(\{x_{j}\},t)|_{\{x_{j}=x_{j}(t)\}}.
\end{equation}
Thus to find this trajectory we need to know either ${\rm
v}_{k}(\{x_{j}\},t)$ or $P(\{x_{j}\},t)$ and $j_{k}(\{
x_{j}\},t)$. In previous literature the method is normally to find
$P(\{x_{j}\},t)$ and $j_{k}(\{x_{j}\},t)$. Here, however, we
propose that ${\rm v}_{k}(\{x_{j}\},t)$ is given by
\begin{equation} \label{VelocityField}
{\rm v}_{k}(\{x_{j}\},t)=\frac{{\rm Re}[\bra{\Psi(t)}\{
x_{j}\}\rangle\bra{\{ x_{j}\}}\hat{\rm
v}_{k}(t)\ket{\Psi(t)}]}{\bra{\Psi(t)}\{
x_{j}\}\rangle\langle\{x_{j}\}\ket{\Psi(t)}},
\end{equation} where $\hat{\rm
v}_{k}(t)$ is called the velocity operator. It is given by
\begin{equation} \label{VelocityOperator}
\hat{\rm v}_{k}(t)=-\frac{i}{\hbar}[\hat{X}_{k},\hat{V}_{\rm
int}(t)].
\end{equation} Note this is only valid for Hamiltonians with terms which are at most quadratic in $\hat{Y}$
(this is proven in Ref. \cite{GamWis03c}). This is not a problem
as in nature all fundamental Hamiltonians are of this form.

Up until now we have not said anything about combined systems. If
we now include an extra system, the system of interest, but only
calculate trajectories for the bath positions, then the following
differs slightly from the conventional de Broglie-Bohm theory.
Defining the system state as
\begin{equation}\label{ConditionedB}
\ket{\psi_{\{x_{j}\}}(t)}=\frac{\bra{\{x_{j}\}}\Psi(t)\rangle}{\sqrt{\langle\Psi(t)\ket{\{x_{j}\}}
\bra{\{x_{j}\}}\Psi(t)\rangle}},
\end{equation}
the velocity field becomes
\begin{equation} \label{vfield}
{\rm v}_{k}(\{x_{j}\},t)={\rm
Re}[\bra{\psi_{\{x_{j}\}}(t)}\overrightarrow{\hat{\rm
v}_{k}(\{x_{j}\},t)}\ket{\psi_{\{x_{j}\}}(t)}],
\end{equation} where
\begin{equation}
\overrightarrow{\hat{\rm
v}_{k}(\{x_{j}\},t)}\ket{\psi_{\{x_{j}\}}(t)}=\frac{\bra{\{x_{j}\}}\hat{\rm
v}_{k}(t)\ket{\Psi(t)}}{\sqrt{\langle\Psi(t)\ket{\{x_{j}\}}
\bra{\{x_{j}\}}\Psi(t)\rangle}}.
\end{equation} Note the arrow defines the direction of operation
[$\overrightarrow{\hat{\rm v}_{k}(\{x_{j}\},t)}$ will contain
partial derivatives the act in the direction of the arrow].
Substituting the velocity field [\erf{vfield}] into
\erf{VelocityA} results in following differential equation for the
bath positions
\begin{equation}\label{VelocityB}
d_{t}{x}_{k}(t)={\rm
Re}[\bra{\psi_{\{x_{j}(t)\}}(t)}\overrightarrow{\hat{
v}_{k}(\{x_{j}(t)\},t)}\ket{\psi_{\{x_{j}(t)\}}(t)}],
\end{equation} where
$\ket{\psi_{\{x_{j}(t)\}}(t)}=\ket{\psi_{\{x_{j}\}}(t)}|_{\{x_{j}=x_{j}(t)\}}$.

We now make the link that \erf{ConditionedB} is mathematically
equivalent to \erf{Conditioned} even though they have different
 physical interpretations. Since in Sec. \ref{sect: QMT} we showed that the
position non-Markovian SSE is derived from this equation, then the
time derivative of \erf{ConditionedB}  must also give the correct
non-Markovian SSE. Thus under this hidden variable interpretation
the solution of the position non-Markovian SSE is a real entity
which exists for all time and guides the trajectories for the bath
positions $\{x_{k}(t)\}$. It gives the `conditioned' system state
at all times, as under the de Broglie-Bohm interpretation the bath
positions exist even when we do not measure the bath.

To show that \erf{VelocityB} does give the same trajectories as
used in the non-Markovian SSE derived from QMT, we used the above
definition for $\hat{\rm v}_{k}(t)$ and apply the Hamiltonians
defined in Sec. \ref{sect: General Dyn} to them. This gives
\begin{equation}
  \hat{\rm v}_{k}(t)=[g_{k}^{*}e^{i\Omega_{k}t}\hat{L}+g_{k}e^{-i\Omega_{k}t}
\hat{L}\dg]/{\sqrt{2}},
\end{equation} which results in a velocity field of the form
\begin{eqnarray}
  {\rm v}_{k}(\{x_{j}\},t)&=&[g_{k}^{*}e^{i\Omega_{k}t}\bra{\psi_{\{x_{j}\}}(t)}
  \hat{L}\ket{\psi_{\{x_{j}\}}(t)}\nl+g_{k}e^{-i\Omega_{k}t}
\bra{\psi_{\{x_{j}\}}(t)}\hat{L}\dg\ket{\psi_{\{x_{j}\}}(t)}]/{\sqrt{2}}.\nl
\end{eqnarray} Thus the actual trajectories are found from
\begin{equation}\label{SDEB}
  d_{t} x_{k}(t)={[\langle\hat{L}\rangle_{t}g_{k}^{*}
  e^{i\Omega_{k}t} +\langle\hat{L}\dg\rangle_{t} g_{k} e^{-i\Omega_{k}
  t}]}/{\sqrt{2}}
\end{equation}
where $\langle
\hat{L}\rangle_{t}=\bra{{\psi}_{\{x_{j}(t)\}}(t)}\hat{L}\ket{{\psi}_{\{x_{j}(t)\}}(t)}$,
which is the same as \erf{SDE}.

\section{DISCUSSION AND CONCLUSION}

In this paper we have firstly extended the theory of non-Markovian
SSE to include the position unraveling. In Sec. \ref{sect: simple}
we applied this theory to a simple system, a TLA coupled linearly
to a single mode, to show that the position non-Markovian SSE does
correctly average to give the reduced state. Although this
unraveling does not have a well defined Markovian limit, it is
still interesting to look at as it allows us to investigate
interpretational questions in quantum mechanics. This is because
this unraveling can be interpreted under either the orthodox
interpretation or by a non-local hidden variable theory which is
similar to the de Broglie-Bohm theory.

Under the orthodox theory the solution of the non-Markovian SSE at
time $t$ is the state the system will be in if a measurement was
performed on the bath at time $t$ and yielded results $\{
x_{k}(t)\}$. Thus the non-Markovian SSE is simply a numerical tool
for calculating the correct conditioned system state. In the de
Broglie-Bohm hidden variable interpretation the non-Markovian SSE
is an evolution equation for the system state (a real state) which
guides the trajectories for the baths individual positions
$\{x_{k}(t)\}$. Unlike the orthodox theory, these variables exist
even when the bath is not measured.  We conclude that if a
non-trivial (non-numerical tool) interpretation is to be given to
this non-Markovian SSE then we must consider the de Broglie-Bohm
hidden variable interpretation of quantum mechanics.

\acknowledgments

This work was supported by the Australian Research Council.


\end{document}